\documentclass[floatfix,twocolumn,aps,prb,showpacs,superscriptaddress]{revtex4}
\usepackage{amssymb}
\usepackage{graphicx}
\usepackage{dcolumn}
\usepackage{bm}

\newcommand{\beq}{\begin{eqnarray}}
\newcommand{\eeq}{\end{eqnarray}
}\def\tm{T_{-1}}
\def\tone{T_1}
\def\tzero{T_0}

\begin{document}

\title{Higher order effective low-energy theories}
\author{A. L. Chernyshev}
\affiliation{Department of Physics, University of California, Irvine, California 92697}
\author{D. Galanakis}
\affiliation{Loomis Laboratory of Physics, University of Illinois at
  Urbana-Champaign, 1110 W.Green St., Urbana, IL., 61801-3080}
\author{P. Phillips}
\affiliation{Loomis Laboratory of Physics, University of Illinois at
  Urbana-Champaign, 1110 W.Green St., Urbana, IL., 61801-3080}
\author{A. V. Rozhkov}
\affiliation{Department of Physics, University of California, Irvine, California 92697}
\author{A.-M. S. Tremblay}
\affiliation{Department of Physics, Yale University, P.O. Box 208120, New Haven, CT
06520-8120}
\affiliation{D\'{e}partement de physique and Regroupement qu\'{e}b\'{e}cois sur les mat%
\'{e}riaux de pointe, Universit\'{e} de Sherbrooke, Sherbrooke, Qu\'{e}bec,
J1K 2R1, Canada}
\date{\today }

\begin{abstract}
Three well-known perturbative approaches to deriving low-energy effective
theories, the degenerate Brillouin-Wigner perturbation theory (projection
method), the canonical transformation, and the resolvent methods are
compared. We use the Hubbard model as an example to show how, to fourth
order in hopping $t$, all methods lead to the same effective theory, namely
the $t$-$J$ model with ring exchange and various correlated hoppings. We
emphasize subtle technical difficulties that make such a derivation less
trivial to carry out for orders higher than second. We also show that in
higher orders, different approaches can lead to seemingly different forms for
the low-energy Hamiltonian. All of these forms are equivalent since they are
connected by an additional unitary transformation whose generator is given
explicitly. The importance of transforming the operators is emphasized and
the equivalence of their transformed structure within the different
approaches is also demonstrated.
\end{abstract}

\pacs{71.10.-w, 71.10.Fd, 74.72.-h, 75.10.Jm}
\maketitle


\section{Introduction}

There is a significant recent interest in higher-order corrections to
effective low-energy theories for a broad range of strongly-correlated
electronic problems.\cite{Hemerle:2003,coldea,Lannert,Sandvik} For example,
the effective low-energy Hamiltonians of the Hubbard-like models contain the
so-called ring-exchange terms.\cite{Girvin:1989,Takahashi:1977} These terms
can alter the basic properties of excitations,\cite{Lannert} or shift the
balance towards a new ground state.\cite{Sandvik} In addition, an
accurate description 
of the experimentally observed\cite{Experiments} spectral weight
transfer over the Mott scale also necessitates high order corrections
in the hopping energy.\cite{Eskes:1994,Harris:1967} Although several
methods exist to derive higher-order low-energy theories, the unicity of the
low-energy effective theory may not be obvious.  In fact, technical
subtleties appear, even in a straightforward application of
Brillouin-Wigner perturbation theory, that may lead to ambiguous
results\cite{Su} beyond second order in the hopping.    Further,
unlike traditional applications of degenerate perturbation theory in
which the eigenstates of the projected Hamiltonian also diagonalize
the perturbation, such is not the case here.\cite{pgs} The unique
difficulty that arises with Hubbard-like models is that part of the
kinetic energy perturbation leaves the number of doubly occupied sites
unchanged and hence must still be iteratively diagonalized in the
low-energy subspace.

Low-energy effective theories play a crucial role in essentially all fields
of physics. When there are large energy scales that are well separated from
the low-energy sector, low-energy effective theories offer the enormous
advantage of being formulated in an exponentially smaller Hilbert space. The
price for this simplification is that both the Hamiltonian \textit{and} the
operators take a form that is more complicated than that of the original
theory. In general, this form involves $n$-body interactions and operators
that have a finite extent in space, even if the original theory was local.
For example, making use of the large energy of virtual electron-positron
pairs, Foldy and Wouthuysen derived non-relativistic quantum mechanics from
the Dirac equation.\cite{Foldy:1950} The local gauge coupling to the
electromagnetic potential in Dirac theory yielded three interactions,
Zeeman, spin-orbit, and Darwin terms, that are all non-local since they
involve derivatives of the electromagnetic potentials. This was done with
the help of a canonical transformation\cite{Foldy:1950} but the
Brillouin-Wigner type methods lead to the same result.\cite{Sakurai:1967} In
the case of the Hubbard model, when the interaction $U$ is large compared
with hopping $t$, the second-order effective Hamiltonian in the
singly-occupied subspace is the $t$-$J$ model (including correlated hopping)
where the spin-spin interaction is non-local in contrast to the original
local Hubbard interaction. In higher order, $n$-body interactions, such as
ring-exchange terms, also appear.\cite{Girvin:1989} The appearance of $n$%
-body interactions and of non-local terms in effective low-energy theories
is familiar in the Wilsonian renormalization group context.\cite{Wilson:1975}

There are several ways to obtain an effective low-energy Hamiltonian, \cite%
{Klein:1974} two of which are widely used in condensed matter physics. The
first one is the canonical transformation (CT) method, mentioned above, \cite%
{Foldy:1950} which is based on original ideas of Van Vleck. \cite%
{VanVleck:1929} The derivation of the $t-J$ model with ring exchange\cite%
{Takahashi:1977,Girvin:1989} from the Hubbard model and the derivation of
the Kondo model using the Schrieffer-Wolff transformation\cite%
{SchriefferWolf} from the Anderson model, are well-known examples where the
CT method was applied. A modification of that method, the continuous CT \cite%
{Wegner:1994,Uhrig:2000} and its predecessor, \cite{Safonov} has also found
recent applications in the derivation of flow equations. Another equally
popular method is the projection method, or degenerate Brillouin-Wigner (BW)
perturbation theory. The projection method, based on ideas of Kato, \cite%
{Kato:1949} offers an alternative route to the derivation of, for example,
the $t-J$ model with correlated hopping.\cite{Anderson:1959,Fazekas}
It has been also used recently 
in the derivation of an effective Hamiltonian for the pyrochlores.\cite%
{Hemerle:2003} The resolvent method, based on a procedure known as Lowdin
downfolding,\cite{Lowdin} is yet another method that is available.

It is well known that the BW and the CT methods give equivalent results at
low (second) order, as exemplified by the $t-J$ model with correlated
hopping. Although the
equivalence of these approaches in higher order might be not obvious, one
intuitively expects that they are. This has been supported by Klein
\cite{Klein:1974} 
who has given a formal proof of the equivalence between various forms of
degenerate perturbation theory. Generally, in the presence of a well defined
small parameter, the perturbative expansion in powers of such a parameter
should not depend on the method employed. In other words, an effective
low-energy theory can be presented as an $n$-th order power series in the
inverse of some large energy scale ($t/U$ in the Hubbard model), and all
methods are expected to yield equivalent forms of the theory up to that same
order $n$.

In this paper we show in some detail how CT, resolvent and BW methods can be
applied to the Hubbard model to obtain equivalent effective
low-energy Hamiltonians up to fourth order, that is $O(t^{4}/U^{3})$.  By equivalence, we mean that all three methods yield Hamiltonians that are related via a unitary transformation.  Although,
in order to be specific, we work with the Hubbard model, it will be clear
that the procedure can be trivially extended to other models, including
models that involve expansion of pure spin models about the Ising limit.
\cite{Lannert} Several new issues appear in deriving higher-order low-energy
effective theories. First, one may find amusing that the low-energy
Hamiltonians obtained from three different CT methods, one of
Refs. \onlinecite{Girvin:1989,Takahashi:1977}, one of
Ref. \onlinecite{Stein}, and the one introduced in this work, appear
to be \textit{different} in each case. However, we show that they are all
connected by an additional unitary transformation that leaves the
block-diagonal form invariant.\cite{Stein} That is, this unitary
transformation converts these different Hamiltonians one into the
other. Second, 
in the case of the BW method one should be careful in dealing with: \textit{%
(i)}\ the orthonormalization of the projected eigenstates, since the latter
are not necessarily orthogonal even if the initial basis is, \textit{(ii)}\
the energy dependence of the expansion, since the energy should be evaluated
iteratively using results from previous steps, in the spirit of Rayleigh-Schr%
\"{o}dinger non-degenerate perturbation theory. The final low-energy theory
again appears to be different from that obtained from the CT approach
of Ref. \onlinecite{Girvin:1989},
but a unitary transformation within the low-energy subspace shows their
equivalence. Third, in the so-called resolvent method, which is similar to the
BW approach but works more directly with the Hamiltonian matrix rather than
with the eigenstates, one needs to perform the orthonormalization of
the eigenstates in an iterative
procedure. For the  derivation of the $t$-$J$ model  with correlated
hopping within this method see, e.g. 
Ref. \onlinecite{auerbach}. The fourth-order low-energy theory obtained with
this method is also unitarily equivalent to the results of the BW and CT
approaches.

Finally, as in Ref. \onlinecite{Eskes:1994}, we emphasize that in
order to compute correlation functions or spectral weight within the
low-energy theory it is important to
transform the operators corresponding to observables.
The omission of such a transformation (see, for example,
the recent work using the BW method for the pyrochlores \cite%
{Hemerle:2003,NoteFisher}) has to be explicitly addressed. We also demonstrate the
equivalence of the transformed structure of the operators within the
different methods.

Our paper is organized as follows. Sec. \ref{Model} introduces the model and
notations. Sec. \ref{CT} describes CT methods and introduces the unitary
transformation that allows us to demonstrate the equivalence of the various
low-energy effective theories. Sec. \ref{DBW} is devoted to the BW method,
 Sec. \ref{resolvent} describes the resolvent method and finally
 Sec. \ref{bigp} presents a generalized transformation from which all
 perturbative results can be derived. In Sec. \ref{Discussion} 
we give a brief
 discussion of relevant experiments. We conclude 
with Sec. \ref{Conclusions}. Appendix \ref{app_A} contains miscellaneous
results.


\section{Model}

\label{Model}

We consider the Hubbard model, conveniently written in the
form \cite{Harris:1967}
\begin{equation}
H=T_{0}+T_{1}+T_{-1}+V,  \label{H}
\end{equation}%
where the \emph{O}$(t)$ kinetic energy operator in second quantized form has
been divided into three terms by using projection operators. The first
operator, $T_{0}$, includes the projection operators that ensure that the
number of doubly-occupied sites does not change because of hopping. It
is precisely this term that complicates the application of traditional
degenerate perturbation theory to the Hubbard model.  The 
projection operators included in the kinetic energy operators $T_{1}$ and $%
T_{-1}$ make sure that these operators increase or decrease the number
of doubly-occupied sites by $1$, respectively. All of these
terms are proportional to the hopping matrix element $t$. Note that
generally speaking, there can be hopping matrix elements to arbitrary
neighbors but we take all these terms to be of the same order in the
expansion parameter $t$. The Hubbard on-site repulsion, written as $V$,
is proportional to $U$.

These notations implicitly use the classification of the Hilbert space in
subspaces with different numbers of doubly-occupied sites. Namely, every
eigenfunction of the Hubbard Hamiltonian (\ref{H}) can be split in the
series of orthogonal pieces
\begin{equation}
|\psi \rangle =\sum_{m=0}^{\infty }|\psi _{m}\rangle =|\psi _{0}\rangle
+|\psi _{1}\rangle +|\psi _{2}\rangle +\dots \ ,  \label{expansion_psi_m}
\end{equation}%
where the subscript enumerates the states with $m=0,1,2,\dots $
doubly-occupied sites. Evidently, the operator $V$ is diagonal in this basis
and has the eigenvalue $mU$ in a state containing $m$ doubly-occupied sites.
For the rest of the paper we set $U=1$, and consider $t$ itself as a small
parameter. When a clarification is needed we will restore the actual $t/U$
dependence of the expression.

At $t=0$, the solution of the eigenstate equation is simply a set of highly
degenerate states separated by energy $U$. It is assumed that $U$ is much
larger than the bandwidth (of order $t$) so that, with $t$ finite,
states cluster around the values $mU$ and are separated from each other by a
Mott gap where no states occur. In other words, we assume that switching on $%
t$ does not lead to a crossing of levels between the lowest energy manifold,
$mU$ with $m=0$, and all other manifolds with $m> 0$.

Evidently, many Hamiltonians can be cast in the form of Eq. (\ref{H}). It
suffices to have a large term in the Hamiltonian that can be easily
diagonalized and leads to states that are separated by some large energy
scale $U$. Other terms in the Hamiltonian that couple these original states
and are proportional to some small parameter can be denoted by
$T_{\Delta m}$ where
$\Delta m$ indicates that the operator couples states separated by $\Delta mU$. Although
we consider only $\Delta m=-1,0,+1,$ it will be obvious how to generalize our proof
to more general values of $\Delta m$.


\section{Canonical Transformations}

\label{CT} \setcounter{subsubsection}{0} 

The CT  method is probably the most commonly used
method to find low-energy effective theories. We briefly recall the known
results for the problem at hand and then move to another version of the
CT approach that gives a result that is unitarily
related to the first one.

\subsubsection{Method 1}

To derive the higher-order effective Hamiltonian for the Hubbard model, the
CT method was applied in Refs. \onlinecite{Takahashi:1977,Girvin:1989}.
There the $t^{5}$- and $t^{6}$-order Hamiltonians were obtained for the
half-filled case and $t^{4}$-order Hamiltonian was found for an arbitrary
doping. We will simply repeat the basic idea and the results obtained in
Ref. \onlinecite{Girvin:1989}. The effective Hamiltonian was obtained from
the Campbell-Baker-Hausdorff expression
\begin{equation}
\mathcal{H}_{eff}^{CT_{1}}=e^{-S}He^{S}=H+[H,S]+\frac{1}{2!}[[H,S],S]+\dots
\ ,  \label{H_can}
\end{equation}%
where the generator of the transformation $S$ is truncated as
\[
S=S_{1}+S_{2}+S_{3}+\dots
\]%
where $S_n\propto t^n$. The role of each
$S_{n}$ in this series is to eliminate the corresponding $t^{n}$-order
off-diagonal terms in the Hamiltonian in Eq. (\ref{H_can}), which change the
number of doubly-occupied sites. By assumption however, $S$ does not contain
terms that preserve the number of doubly-occupied sites. The remaining
freedom to perform a unitary transformation within the singly-occupied
subspace will be discussed later in this section.

Given the explicit form of the Hubbard Hamiltonian, Eq. (\ref{H}), one
readily finds that
\begin{equation}
S_{1}=T_{-1}-T_{1}\ .  \label{S1}
\end{equation}%
Using $S=S_{1}$ and keeping terms up to $O(t^{2})$ in Eq. (\ref{H_can}) the $%
t-J$ model with correlated hopping 
is obtained. To derive the higher-order Hamiltonians we have to
truncate $S$ at higher order and determine the operator expression of $S_{n}$%
's required to eliminate the off-diagonal terms of the $t^{n}$ order. These
off-diagonal terms are generated by the commutators of $H$ with $%
S_{n^{\prime }}$'s in the previous orders. Generally, one needs $n-1$ terms
in $S$ to obtain the theory valid to the order $t^{n}$. We list here the
generator $S_{2}$ for completeness
\begin{equation}
S_{2}=T_{0}T_{-1}-T_{1}T_{0}-T_{-1}T_{0}+T_{0}T_{1}\ ,  \label{S2}
\end{equation}%
and simply reproduce the result of the procedure described above carried out
in Ref. \onlinecite{Girvin:1989} to the $t^{4}$-order,
\begin{widetext}
\begin{eqnarray}
\mathcal{H}^{(4),CT_1}_{eff}&=&V+T_0-T_{-1}T_1
+T_{-1}T_0T_1-\frac12(T_{-1}T_1T_0+T_0T_{-1}T_1)+
(\tm\tone)^2-\frac12\tm^2\tone^2  \nonumber \\
&&-T_{-1}\tzero^2T_1+T_{-1}T_0T_1T_0+T_0T_{-1}T_0T_1
-\frac12(T_{-1}T_1\tzero^2+\tzero^2T_{-1}T_1) \ .  \label{AM}
\end{eqnarray}
\end{widetext}
The value of $V$ is taken to be zero since we are in the singly-occupied
subspace. In deriving this expression the identity $T_{-1}|\psi _{0}\rangle
=0$, where $|\psi _{0}\rangle $ is any of the singly-occupied states, is used.

We would like to add an interesting technical detail to the discussion of
this method. The $t^{3}$-term in the generator, $S_{3}$, although necessary
to eliminate the $t^{3}$-order off-diagonal terms in Eq. (\ref{H_can}), does
not contribute to $\mathcal{H}_{eff}^{(4),CT_{1}}$ explicitly. That is, the 4%
$th$ order effective Hamiltonian of Eq. (\ref{AM}) can be obtained using $%
S=S_{1}+S_{2}$ only, simply neglecting the remaining off-diagonal terms.
Similarly, the diagonal terms in the 3$rd$-order Hamiltonian $\mathcal{H}%
_{eff}^{(3)}$ are all generated by $S_{1}$ alone. This is because the
original Hubbard Hamiltonian does not contain any \textquotedblleft
bare\textquotedblright\ off-diagonal terms of order higher than $t$. All
such higher-order off-diagonal terms are the result of commutations of $T$%
-terms. This shows some additional internal structure of the model. Since we
will need the generator $S_{3}$ for the discussion of the transformed operators
but it was not written out explicitly in Ref. \onlinecite{Girvin:1989}, we
present it in Appendix \ref{app_A}, Eq. (\ref{S3}).

The expression for the $\mathcal{H}_{eff}^{(4),CT_{1}}$ in Eq. (\ref{AM})
seems to have all possible combinations of $T_{0}$, $T_{-1}$, and $T_{1}$,
except for one \textquotedblleft missing term\textquotedblright : $%
T_{0}T_{-1}T_{1}T_{0}$.
Although there is no general principle which would require presence
of such a term in the
effective Hamiltonian, its absence makes one curious about its whereabouts.
The fate of this term will be clarified in Sec. \ref{addCT}.

\subsubsection{Method 2}

In the same spirit, a different way to formulate an effective theory using
the CT approach is to apply consecutive unitary transformations
\[
\mathcal{H}_{eff}^{CT_{2}}=\dots e^{-\widetilde{S}_{3}}e^{-\widetilde{S}%
_{2}}e^{-\widetilde{S}_{1}}\ H\ e^{\widetilde{S}_{1}}e^{\widetilde{S}_{2}}e^{%
\widetilde{S}_{3}}\dots \ .
\]%
We call this the \textquotedblleft consecutive CTs\textquotedblright\
approach. The idea for each $\widetilde{S}_{n}$ is to eliminate the
off-diagonal terms of the $n$th order remaining from the previous, $n-1$
order CT. In each of the transformations the expansion formula Eq. (\ref%
{H_can}) is applied and all terms up to a desired order in $t$ are kept.
Thus, after $\widetilde{U}_{1}=e^{\widetilde{S}_{1}}$ is applied to the
original Hamiltonian the off-diagonal terms of order $O(t)$ are eliminated
and show up only in $O(t^{2})$. The next transformation moves the
off-diagonal terms to $O(t^{3})$, and so on. Generally, the generators $%
\widetilde{S}_{n}$ in this approach are different from the ones in the
previous approach, that is $\widetilde{S}_{n}\neq S_{n}$. However, one can
check that for the Hubbard model generators $\widetilde{S}_{1}=S_1$ and $%
\widetilde{S}_{2}=S_2$, Eqs. (\ref{S1}) and (\ref{S2}).
Note that $\widetilde{S}_{3}$ is indeed different from $S_{3}$.

We would like to remark that the derivation of the higher-order effective
Hamiltonian within this \textquotedblleft consecutive CTs\textquotedblright\
approach is more straightforward than the approach of Ref. %
\onlinecite{Girvin:1989}. Also, the absence of the $\widetilde{S}_{3}$
contribution to $\mathcal{H}_{eff}^{(4)}$ is much less enigmatic here.
Namely, since $\widetilde{S}_{3}$ is the off-diagonal order $t^{3}$
operator, its only commutator which can give $O(t^{4})$ contribution to the
Hamiltonian is with the off-diagonal operator of the order $t$. However,
after $e^{\widetilde{S}_{1}}$ is applied the only operator of order $t$
remaining in the transformed Hamiltonian is the {\it diagonal} operator $T_{0}$.
This removes the need to know the explicit form of $\widetilde{S}_{3}$,
although it is formally still necessary to eliminate the off-diagonal $t^{3}$%
-order terms in $\mathcal{H}_{eff}^{(4)}$.

Surprisingly, the final 4$th$ order result of the \textquotedblleft
consecutive CTs\textquotedblright\ approach is \textit{different} from $%
\mathcal{H}_{eff}^{(4),CT_{1}}$ Eq. (\ref{AM}):
\begin{eqnarray}
\mathcal{H}_{eff}^{(4),CT_{2}} &=&\mathcal{H}_{eff}^{(4),CT_{1}}+\frac{1}{2}%
(\tzero^2T_{-1}T_{1}  \label{H_CUT} \\
&&+T_{-1}T_{1}\tzero^2)-T_{0}T_{-1}T_{1}T_{0}\ .  \nonumber
\end{eqnarray}%
The difference concerns the above-mentioned \textquotedblleft missing
term\textquotedblright\ $T_{0}T_{-1}T_{1}T_{0}$ and the terms $%
\tzero^2T_{-1}T_{1}$ and $T_{-1}T_{1}\tzero^2$.
In fact, in this version of the effective theory the original
\textquotedblleft missing term\textquotedblright\ is found, but the two
analogous terms are missing. This ``mystery'' is unveiled below.


\subsubsection{Additional unitary transformation}

\label{addCT} 
Let us first make the following observation. One can consider the following
unitary transformation:
\begin{equation}
\mathcal{H}_{eff}^{\prime }=e^{-S_{0}}\mathcal{H}_{eff}^{CT_{1}}e^{S_{0}}\ ,
\label{Canonical_T}
\end{equation}%
with the generator
\begin{equation}
S_{0}=\gamma (T_{0}T_{-1}T_{1}-T_{-1}T_{1}T_{0})\ ,  \label{S0}
\end{equation}%
where $\gamma $ is a real number and plays the role of an \textquotedblleft
angle of rotation\textquotedblright . Note that such a generator: \textit{%
(i)} is explicitly anti-hermitian, \textit{(ii)} is diagonal (does not change the
number of doubly-occupied sites), \textit{(iii)} is $O(t^{3})$, and
\textit{(iv)} is real. The operator $S_0$ is the lowest order operator
satisfying \textit{(i)}-\textit{(iv)}
which one can construct using $T_{0}$ and $T_{\pm 1}$. It is also the only
operator of such kind in the $O(t^{3})$ order. Therefore, the only
contribution from such a transformation to $\mathcal{H}_{eff}^{\prime }$
will be from the $[T_{0},S_{0}]$ commutator and it will generate additional
terms of $t^{4}$-order of the form,
\begin{eqnarray}
\delta \mathcal{H}_{eff}=\gamma
\big(\tzero^2\tm\tone+\tm\tone\tzero^2-2\tzero\tm\tone\tzero\big) . 
\label{Canonical_T_explicit}
\end{eqnarray}%
As we will see, the multiplicity of Hamiltonians that arise once the
high energy scale is eliminated all differ by the terms appearing in
$\delta \mathcal{H}_{eff}$. 
Choosing the \textquotedblleft angle of rotation\textquotedblright\ $\gamma
=1/2$ and applying the transformation $S_{0}$ to the $\mathcal{H}%
_{eff}^{(4),CT_{1}}$, Eq. (\ref{AM}), one readily obtains $\mathcal{H}%
_{eff}^{(4),CT_{2}}$, Eq. (\ref{H_CUT}). As a result $T_{-1}T_{1}\tzero^2+%
\mbox{H.c.}$ are replaced by the \textquotedblleft missing
term\textquotedblright\ $T_{0}T_{-1}T_{1}T_{0}$. Clearly, different choices
of the \textquotedblleft angle of rotation\textquotedblright\ will give
different fractions of those terms in the result. In fact, a recent study,
Ref.
\onlinecite{Stein}%
, used a continuous CT approach to the Hubbard model and obtained an
effective Hamiltonian which would be equivalent to the choice $\gamma =1/4$
in Eq. (\ref{S0}).

Thus, the 4$th$-order effective Hamiltonian for the Hubbard model can take
an infinite number of unitary equivalent \textit{forms}, all connected by
the transformation in Eqs. (\ref{Canonical_T}), (\ref{S0}). All these models
possess the same energy spectrum and correlation functions and thus are
equivalent. From this point of view, the reader should not be surprised
when, in the next section, we find that the BW method gives a result that is
different from Eq. (\ref{AM}).

Although the unitary equivalence of the models is a rather natural property,
it is certainly unfamiliar in lower-order effective theories. Furthermore,
such a unitary equivalence should be common to all higher-order theories. As
the order of the perturbation theory is increased the number of
block-diagonal, anti-hermitian operators one can construct will also grow,
providing one with a broader variety of unitarily equivalent forms of the
effective Hamiltonian and corresponding operators.%

\subsubsection{Operators}

It is important to note that in
a low-energy effective theory all operators should be transformed along
with the Hamiltonian. Then the expectation values of the
observables can be calculated in the singly-occupied manifold. The
transformation is different depending on which canonical transformation
method is used. Using the first CT above, the standard expression for the
transformation is
\[
\widetilde{O}=e^{-S}Oe^{S}=O+[O,S]+\frac{1}{2!}[[H,O],O]+\dots \ .
\]%
Again, we consider as an example the operator $O_{1}$, which increases the
number of doubly-occupied sites by one. Using $S=S_{1}+S_{2}$ from Eqs. (\ref%
{S1}) and (\ref{S2}) and utilizing the property $T_{-1}|\psi _{0}\rangle =0$
we obtain, to the order $t^{2}$
\[
\widetilde{O}_{1}=-T_{-1}O_{1}+(T_{-1}T_{0}-T_{0}T_{-1})O_{1},
\]%
which coincides with the expression we will obtain with the BW method, Eq. (%
\ref{O_2}). To obtain the next-order expression for the transformed operator
one needs to know an explicit expression for the generator $S_{3}$ (see
Appendix \ref{app_A}, Eq. (\ref{S3})). Using it, some algebra reveals that,
to order $t^{3}$
\begin{eqnarray}
\widetilde{O}_{1} &=&-T_{-1}O_{1}+(T_{-1}T_{0}-T_{0}T_{-1})O_{1}  \nonumber
\\
&&+\frac{3}{2}T_{-1}T_{1}T_{-1}O_{1}-\frac{1}{2}\tm^2T_{1}O_{1} \\
&&-\tzero^2T_{-1}O_{1}-T_{-1}\tzero^2O_{1}+2T_{0}T_{-1}T_{0}O_{1}
\nonumber \\
&&+\frac{1}{2}T_{-1}O_{1}T_{-1}T_{1}-\frac{1}{2}\tm^2O_{1}T_{1},
\nonumber  \label{O_CT3}
\end{eqnarray}%
which should be compared with the result of the BW method given in Appendix %
\ref{app_A}, Eq. (\ref{O_3A}).

\section{Brillouin-Wigner method}

\label{DBW}\setcounter{subsubsection}{0}

We proceed to show, up to order $O(t^{4})$ ($=O(t^{4}/U^{3})$) for the
Hamiltonian, that degenerate BW perturbation theory can be organized in the
spirit of the Rayleigh-Schr\"{o}dinger (RS) perturbation theory to lead to
the same low-energy effective theory for Hamiltonian and operators as
the consecutive CT
method. We will take a detailed approach that shows all the subtleties.

Let us consider one eigenstate $|\psi \rangle $ of the full Hamiltonian (\ref%
{H}) with the eigenvalue $E$. It obeys the Schr\"{o}dinger equation
\begin{equation}
\left( E-T_{0}-V-T_{1}-T_{-1}\right) |\psi \rangle =0.\
\label{Schroedinger}
\end{equation}%
Although we have not explicitly written quantum numbers for $E$ and $|\psi
\rangle $, we have to remember that we have a matrix equation with many
eigenvalues and corresponding eigenstates. We look for the effective theory
that describes the states that evolve from the lowest energy sector, $m=0,$
taking into account virtual excitations into $m> 0$ states
perturbatively. One can also write effective theories that are valid for any
of the subspaces \cite{Cohen} with $m> 0$.


\subsubsection{BW expression for $\vert\protect\psi\rangle$}

We would like to rewrite the eigenstate $|\psi \rangle $ in a way that will
allow us to take into account higher-energy sectors with $m~\geq ~1$ through
an iterative procedure. Let $Q$ be a projection operator that removes all
components of $|\psi \rangle $ that are in the $m=0$ (singly-occupied)
subspace. We have $\left[ Q,E-T_{0}-V\right] =0$ since $T_{0}+V$ does not
change double-occupancy. One can find then an iterative expression for $%
Q|\psi \rangle $ directly from the Schr\"{o}dinger equation Eq. (\ref%
{Schroedinger})
\begin{equation}
Q|\psi \rangle =\frac{1}{E-T_{0}-V}Q\left( T_{1}+T_{-1}\right) |\psi \rangle
.\label{qpsi}
\end{equation}%
Inversion of the operator $E-T_{0}-V$ does not cause any problem when there
is a Mott gap since the denominator has only non-vanishing eigenvalues.
Indeed, $E-T_{0}$ is at most of order of the bandwidth (proportional to $t$)
while the operator $Q$ ensures that the smallest value that $V$ takes is $%
U$. The complete eigenvector $|\psi \rangle $ has components in the $m=0$
subspace $|\psi _{0}\rangle \equiv \left( 1-Q\right) |\psi \rangle $ that we
need to determine. We assume that $|\psi _{0}\rangle $ is a member of an
orthonormal subspace $\langle \psi _{0}^{\prime }|\psi _{0}\rangle =\delta
_{\psi _{0}^{\prime },\psi _{0}}$. The subscript $0$ to a ket means that it
has components only in the $m=0$ subspace. This procedure leads to the
standard BW expression for $|\psi \rangle $ \cite{Negele}
\begin{equation}
|\psi \rangle =|\psi _{0}\rangle +\frac{1}{E-T_{0}-V}Q\left(
T_{1}+T_{-1}\right) |\psi \rangle \;,  \label{Projection}
\end{equation}%
which can be solved perturbatively by iteration. Using $T_{-1}|\psi
_{0}\rangle =0$ and the fact that we cannot come back to the $m=0$ subspace
in any of the intermediate steps $(QT_{-1}T_{1}|\psi _{0}\rangle =0)$ we
find, iterating Eq. (\ref{Projection}) three times,
\begin{widetext}
\begin{eqnarray}
|\psi \rangle =\left[ 1+\frac{1}{E-T_{0}-V}T_{1}+\left(
 \frac{1}{E-T_{0}-V}T_{1}\right)^{2}\right.  &+&\left(
 \frac{1}{E-T_{0}-V}T_{1}\right) ^{3}
\label{BW} \\
&+&\left. \frac{1}{E-T_{0}-V}T_{-1}\left( \frac{1}{E-T_{0}-V}T_{1}\right)
^{2}\right] |\psi _{0}\rangle +\dots \ .  \nonumber
\end{eqnarray}
\end{widetext}
We took into account the projection operators $Q$ so that the above equation
contains only the terms for which $Q$ equals unity. The above equation Eq. (%
\ref{BW}) generates the usual Brillouin-Wigner perturbation theory. One
recognizes that the second term and the last term in Eq. (\ref{BW}) are
components of the eigenvector in the subspace with $m=1$ doubly-occupied
site $|\psi _{1}\rangle $, while the third and the fourth terms are
components with $m=2$ ($|\psi _{2}\rangle $) and $m=3$ ($|\psi _{3}\rangle $%
) doubly-occupied sites, respectively. This form (without $|\psi _{3}\rangle
$) suffices for our derivation of the effective theory to order $\emph{O}%
(t^{4})$. With this effective Hamiltonian, one will be able to find the
component of the eigenstate in the singly occupied subspace $|\psi
_{0}\rangle $. Given $|\psi _{0}\rangle $, all the components of the
eigenvector $|\psi \rangle $ in the $m=1$ and $m=2$ subspaces are already
completely determined by Eq. (\ref{BW}).

The subsequent treatment of Eq. (\ref{BW}) to generate a low-energy theory
is the following. The denominators $(E-T_{0}-V)^{-1}$ are not singular
because they correspond to the energy in the bands with $m> 0$. Using $%
V\gg E-T_{0}$, one has to expand the energy denominators in Eq. (\ref{BW})
to the required order in $t$. Let us list here the results of such an
expansion of Eq. (\ref{BW}) order by order $\left( U=1\right) $. We find
\begin{widetext}
\begin{eqnarray}
|\psi ^{(0)}\rangle  &=&|\psi _{0}^{(0)}\rangle ,  \label{psi0} \\
|\psi ^{(1)}\rangle  &=&\big(1-T_{1}\big)|\psi _{0}^{(1)}\rangle ,
\label{psi1} \\
|\psi ^{(2)}\rangle  &=&\bigg(
1-T_{1}-(E-T_{0})T_{1}\bigg) |\psi_{0}^{(2)}
\rangle ,  \label{psi2} \\
|\psi ^{(3)}\rangle  &=&\left(
1-T_{1}+T_{0}T_{1}-T_{1}E-\frac{1}{2}T_{-1}T_{1}^{2}-T_{1}E^{2}-
\tzero^2T_{1}+2T_{0}T_{1}E\right.
\label{psi3} \\
&&\left.
+\frac{1}{2}\tone^2-\frac{1}{2}T_{1}T_{0}T_{1}+
\frac{3}{4}\tone^2E-\frac{1}{4}T_{0}\tone^2\right)
|\psi _{0}^{(3)}\rangle ,
\nonumber
\end{eqnarray}
\end{widetext}
where the superscript describes the order of approximation, that is $|\psi
_{0}^{(3)}\rangle $ is the component of the third-order eigenstate $|\psi
^{(3)}\rangle $ in the singly-occupied, $m=0$ subspace:
$|\psi_{0}^{(3)}\rangle =(1-Q)|\psi^{(3)}\rangle$.
 Note that, as usual,
the order $t^{n-1}$ in the expansion of the eigenvector $|\psi \rangle $
correspond to the order $t^{n}$ in the matrix elements of the Hamiltonian
and of the corresponding energy. \cite{LL} That is, one computes the
first-order $E^{(1)}$ using zeroth-order eigenstates $|\psi ^{(0)}\rangle $,
the second-order theory $H^{(2)}$ is formulated with the first-order basis $%
|\psi ^{(1)}\rangle $, etc. Thus, for the fourth-order effective Hamiltonian
we will need $|\psi ^{(3)}\rangle $. Another detail concerns the explicit
dependence of $|\psi ^{(2)}\rangle $ and $|\psi ^{(3)}\rangle $ in Eqs. (\ref%
{psi2}), (\ref{psi3}) on the energy $E$. This issue will be resolved later
by evaluating $E$ in iterative manner.

One can see that in all orders the full eigenstate $|\psi \rangle $ is built
from the $m=0$ states $|\psi _{0}\rangle $ by including the off-diagonal
transitions to the upper-band. The zeroth-order approximation $|\psi
^{(0)}\rangle $ Eq. (\ref{psi0}) corresponds to taking only the first term
in Eq. (\ref{BW}) and neglecting all the upper-band excitations. For the
first-order approximation $|\psi ^{(1)}\rangle $, we needed to include the
second term in Eq. (\ref{BW}). Using that $E-T_{0}$ will be of the order of $%
t$ we have simply neglected $E-T_{0}$ in the denominator which yielded the
first-order state, Eq. (\ref{psi1}). The second-order state $|\psi
^{(2)}\rangle $ is obtained from the first three terms in the BW series Eq. (%
\ref{BW}). In the third term we can neglect $E-T_{0}$ again, but the second
term needs to be expanded in $E-T_{0}$. This leads to the second-order state
given in Eq. (\ref{psi2}). A term of the form $\tone^2/2$, which does not
contribute to the effective Hamiltonian to that order, was dropped although
it can appear in certain observables. To obtain the effective Hamiltonian
valid to order $t^{4}$ one has to generate the third-order wave function $%
|\psi ^{(3)}\rangle $ from Eq. (\ref{BW}). This is obtained from the
contributions of all terms in Eq. (\ref{BW}), although the fourth term $%
\propto \tone^3$ can be neglected since it only contributes to the
effective theory to order $t^{6}$. The expansion of the denominators
provides us with the $E$-, and $E^{2}$-dependent terms in the resulting
expression for $|\psi ^{(3)}\rangle $, Eq. (\ref{psi3}). Let us emphasize
here that in the $E$-dependent terms of the second- and third-order
eigenfunctions, one should expand $E$ as well, and keep only the terms to
the required order. Thus, for $|\psi ^{(2)}\rangle $ we will only need the
expression for $E$ that is valid up to order $t$, while for $|\psi
^{(3)}\rangle $ order $t^{2}$ is required.

We also point out that equations (\ref{psi0})-(\ref{psi3}) relate the
\textquotedblleft full\textquotedblright\ eigenstate $|\psi \rangle $ to the
the state in the \textquotedblleft projected\textquotedblright , $m=0$
subspace. Since our goal is to have a low-energy theory which operates with
the projected $|\psi _{0}\rangle $ states only, one should take into account
the fact that although the eigenstates are orthonormal $\langle \psi
^{\prime }|\psi \rangle =\delta _{\psi ^{\prime },\psi }$ this is not true
for the $m~=~0$ component alone, that is $\langle \psi _{0}^{\prime }|\psi
_{0}\rangle \neq \delta _{\psi _{0}^{\prime },\psi _{0}}$. This is evident
from Eq. (\ref{expansion_psi_m}). Therefore, if one wishes to obtain an
effective theory that takes the usual Hermitian form, one needs to
orthonormalize the projected basis to the required order in $t$. This
difficulty will appear in higher order, but let us first reproduce the
well-known second-order results.


\subsubsection{Derivation of the t-J model}

We begin with the zeroth-order approximation $|\psi ^{(0)}\rangle $ given in
Eq. (\ref{psi0}). Since the eigenfunction involves only the $m=0$ subspace,
the diagonal part of the Hamiltonian (\ref{H}) solely contributes to the
eigenvalue equation
\begin{equation}
\langle \psi ^{\prime (0)}|H|\psi ^{(0)}\rangle =\langle \psi _{0}^{\prime
(0)}|T_{0}|\psi _{0}^{(0)}\rangle =E\delta _{\psi _{0}^{\prime },\psi _{0}}\
.  \label{E0}
\end{equation}%
Therefore, $\mathcal{H}_{eff}^{(1)}\equiv T_{0}$ is our effective
Hamiltonian to order $t$. We rewrite Eq. (\ref{E0}), for future reference,
as
\begin{equation}
T_{0}|\psi _{0}^{(0)}\rangle =E|\psi _{0}^{(0)}\rangle \ .  \label{0}
\end{equation}%
To next order, we need the first-order state $|\psi ^{(1)}\rangle $ from Eq.
(\ref{psi1}). One can see that the states $|\psi _{0}^{(1)}\rangle $ in Eq. (%
\ref{psi1}) are still orthonormal to order $t$, that is $\langle \psi
^{\prime (1)}|\psi ^{(1)}\rangle =\langle \psi _{0}^{\prime (1)}|\psi
_{0}^{(1)}\rangle +O(t^{2})=\delta _{\psi ^{\prime },\psi }$. With these
states one obtains
\begin{eqnarray}
\langle \psi ^{\prime (1)}|H|\psi ^{(1)}\rangle  &=&\langle \psi
_{0}^{\prime (1)}|T_{0}-T_{-1}T_{1}|\psi _{0}^{(1)}\rangle   \label{H1} \\
&=&E\langle \psi _{0}^{\prime (1)}|\psi _{0}^{(1)}\rangle +O(t^{3})=E\delta
_{\psi _{0}^{\prime },\psi _{0}}\ ,  \nonumber
\end{eqnarray}%
where we used that $E=O(t)$ and that with the required accuracy we can
neglect $O(t^{3})$ term in the right-hand side and use orthonormality to
replace $\langle \psi _{0}^{\prime (1)}|\psi _{0}^{(1)}\rangle $ by $\delta
_{\psi _{0}^{\prime },\psi _{0}}$. It is clear that the content of the
brackets in Eq. (\ref{H1}) plays the role of an effective second-order
Hamiltonian
\begin{equation}
\mathcal{H}_{eff}^{(2)}\equiv T_{0}-T_{-1}T_{1}\ .  \label{h11}
\end{equation}%
Writing this result in terms of second-quantized operators and recalling
that $J=4t^{2}/U$ in the second term, one recovers the $t-J$ Hamiltonian
with correlated hopping. Again, for future reference, we write: 
\begin{equation}
\big(T_{0}-T_{-1}T_{1}\big)|\psi _{0}^{(1)}\rangle =E|\psi _{0}^{(1)}\rangle
\ .  \label{1}
\end{equation}

Contrary to the non-degenerate perturbation theory, the states in the $m=0$
subspace change as we improve the approximation. There is a link between the
states at various order as will be discussed in more detail in Sec. \ref%
{EigenstatesLink} below.


\subsubsection{Transformation of operators in the t-J model}

Let us pause momentarily to develop transformation rules for the operators
that should be used at this level of approximation. Consider, as an example,
an operator $O_{-1}$ that decreases the number of doubly-occupied states by
one. Naively, one would expect that it has zero expectation in the case of
the $t$-$J$ model that is defined in the singly-occupied subspace. The
correct way to proceed is to notice that the matrix elements of $O_{-1}$ in
the basis of first-order eigenstates Eq. (\ref{psi1}) are given by 
\begin{equation}
\langle \psi ^{\prime (1)}|O_{-1}|\psi ^{(1)}\rangle =\langle \psi
_{0}^{\prime (1)}|-O_{-1}T_{1}|\psi _{0}^{(1)}\rangle \ .
\end{equation}%
Since the above expression is valid for any eigenstate, to first order the
effective operator 
\begin{equation}
O_{-1}^{(1)}=-O_{-1}T_{1}  \label{O-1}
\end{equation}%
should be used to compute any matrix element of the original operator solely
in terms of projected eigenstates $|\psi _{0}\rangle $. Both the Hamiltonian
Eq. (\ref{h11}) and the operator Eq. (\ref{O-1}) coincide with the CT result.


\subsubsection{Third order\label{higher_orders}}

There are new technical issues that appear in orders higher than second. The
first problem is that since $\langle \psi ^{\prime (2)}|\psi ^{(2)}\rangle
=\langle \psi _{0}^{\prime (2)}|1+T_{-1}T_{1}|\psi _{0}^{(2)}\rangle +\emph{O%
}(t^{3})$, the second-order eigenvectors in the singly occupied subspace $%
|\psi _{0}^{(2)}\rangle $ in Eq. (\ref{psi2}) do not form an orthonormal set
to order $t^{2}$. This issue is easily resolved. To obtain an eigenvalue
problem in standard form we define the orthonormal basis $|\varphi
_{0}^{(2)}\rangle $ by 
\begin{equation}
|\psi _{0}^{(2)}\rangle =\left( 1+T_{-1}T_{1}\right) ^{-1/2}|\varphi
_{0}^{(2)}\rangle   \label{phi2}
\end{equation}%
where the square root needs to be expanded to order $t^{2}$ to give 
\begin{equation}
|\psi ^{(2)}\rangle =\left( 1-T_{1}-(E-T_{0})T_{1}-\frac{1}{2}%
T_{-1}T_{1}\right) |\varphi _{0}^{(2)}\rangle .  \label{psi2normalized}
\end{equation}%
This second-order eigenvector $|\psi ^{(2)}\rangle $ should be used to
obtain the third-order effective Hamiltonian. This is where one encounters
the second difficulty. It concerns $E$-dependent terms that have to be
treated carefully.

Suppose $|\psi ^{(2)}\rangle $ in Eq. (\ref{psi2normalized}) is an
eigenstate with energy $E$. Since the order of the $E$-dependent term should
match the $t^{2}$-order of the eigenstate $|\psi ^{(2)}\rangle $, $E$ should
be expanded in $t$. Clearly, we can replace $E$ appearing in Eq. (\ref%
{psi2normalized}) by its first-order term in powers of $t$. To this end we
write
\begin{equation}
E|\varphi _{0}^{(2)}\rangle =\mathcal{H}_{eff}^{(3)}|\varphi
_{0}^{(2)}\rangle \ .  \label{2_0}
\end{equation}%
Since we can anticipate $\mathcal{H}_{eff}^{(3)}$ to have the form
\begin{equation}
\mathcal{H}_{eff}^{(3)}=T_{0}-T_{-1}T_{1}+\emph{O}(t^{3})\ ,  \label{2}
\end{equation}%
we can use $E|\varphi _{0}^{(2)}\rangle =T_{0}|\varphi _{0}^{(2)}\rangle
+O(t^{2})$. This is an iterative procedure and it can always be justified by
the consistency of the result with the initial expectation outlined in Eq. (%
\ref{2}). The replacement $E|\varphi _{0}^{(2)}\rangle =T_{0}|\varphi
_{0}^{(2)}\rangle $ is valid for any of the eigenvectors $|\varphi
_{0}^{(2)}\rangle $ of the effective Hamiltonian. After making this change
in Eq. (\ref{psi2normalized}), the second-order eigenstate reads%
\begin{widetext}
\begin{equation}
|\psi ^{(2)}\rangle =\left( 1-T_{1}-T_{1}T_{0}+T_{0}T_{1}-\frac{1}{2}%
T_{-1}T_{1}\right) |\varphi _{0}^{(2)}\rangle .  \label{Psi2}
\end{equation}%
This procedure allows us to find a Hamiltonian matrix. We will show in Sec. %
\ref{EigenstatesLink} that there is a relation between eigenvectors and
eigenvalues at different orders. With the above state $|\psi ^{(2)}\rangle $%
, the eigenvalue problem takes the form
\begin{equation}
\langle \psi ^{\prime (2)}|H|\psi ^{(2)}\rangle =\langle \varphi
_{0}^{\prime (2)}|T_{0}-T_{-1}T_{1}-\frac{1}{2}%
T_{-1}T_{1}T_{0}+T_{-1}T_{0}T_{1}-\frac{1}{2}T_{0}T_{-1}T_{1}|\varphi
_{0}^{(2)}\rangle =E\delta _{\varphi _{0}^{\prime },\varphi _{0}} \ ,
\label{Eigenequation2}
\end{equation}%
%
%
%
%
which is clearly consistent with our expectation for the eigenvalue $E$
expressed in Eq. (\ref{2}). As in the lower orders, the effective low-energy
Hamiltonian can be directly read off this equation. Thus, up to order $%
O(t^{3})$, the effective Hamiltonian is given by
\begin{equation}
\mathcal{H}_{eff}^{(3)}=T_{0}-T_{-1}T_{1}-\frac{1}{2}%
T_{-1}T_{1}T_{0}+T_{-1}T_{0}T_{1}-\frac{1}{2}T_{0}T_{-1}T_{1}\ ,  \label{Heff3}
\end{equation}%
which agrees with the CT approach to this order (see Sec. \ref{CT}).

The last equation, when taken out of context, may suggest that it contains
terms that are not allowed within BW perturbation theory. Indeed, the
general expression with projected wave function Eq. (\ref{Projection}) shows
that one cannot come back to the singly occupied state in any intermediate
state. The series should contain only \textquotedblleft
proper\textquotedblright\ terms. So, terms like $T_{-1}T_{1}T_{0}$ in the
above expression leave the impression that they should be forbidden since $%
T_{0}$ does not change the double occupancy. However, our derivation clearly
shows that these terms come from expanding $E$ using lower order results.
Hence they are in fact perfectly acceptable.

The matrix elements of any operator to second order
should be computed with the state $|\psi ^{(2)}\rangle $ from Eq. (\ref{Psi2}%
). For the operator $O_{1}$, which increases the number of doubly-occupied
sites by one, we obtain
\begin{equation}
\langle \psi ^{\prime (2)}|O_{1}|\psi ^{(2)}\rangle =\langle \varphi
_{0}^{\prime (2)}|-T_{-1}O_{1}+(T_{-1}T_{0}-T_{0}T_{-1})O_{1}|\varphi
_{0}^{(2)}\rangle\ ,  \label{O_2}
\end{equation}%
that also coincides with the CT result for the transformed operator, as
discussed in Sec. \ref{CT}.

\subsubsection{Fourth order}

To obtain the effective Hamiltonian valid to the order $t^{4}$ one needs
again \textit{(i)} to orthonormalize the states $|\psi _{0}^{(3)}\rangle $,
Eq.~~(\ref{psi3}) and \textit{(ii)} to transform the $E$-dependent terms to
equivalent operator expressions found in the previous steps. The procedure
outlined for the second-order $|\psi ^{(2)}\rangle $ should be followed
here. From the normalization condition of $|\psi ^{(3)}\rangle $ in Eq. (\ref%
{psi3}), one finds that $|\psi _{0}^{(3)}\rangle $ should be orthonormalized
with the help of
\begin{equation}
|\psi _{0}^{(3)}\rangle =\left(
1+T_{-1}T_{1}-2T_{-1}T_{0}T_{1}+2T_{-1}T_{1}E\right) ^{-1/2}|\varphi
_{0}^{(3)}\rangle \ .  \label{psi3_a}
\end{equation}%
The subsequent expansion of the square root to order $t^{3}$ is needed. The
resulting expression for the eigenstate $|\psi ^{(3)}\rangle $ will contain
a term $-T_{1}E$. One has to to replace $E|\varphi _{0}^{(3)}\rangle $ with $%
(T_{0}-T_{-1}T_{1})|\varphi _{0}^{(3)}\rangle $, which gives $E$ to second
order in $t$ when $|\varphi _{0}^{(3)}\rangle $ is an eigenstate. This leads
to the second- and third-order terms.\cite{NoteFisher} In all other $E$%
-dependent terms in Eqs. (\ref{psi3}), (\ref{psi3_a}) we can use $E|\varphi
_{0}^{(3)}\rangle =T_{0}|\varphi _{0}^{(3)}\rangle $ as before since this
already produces the terms of order $t^{3}$. Straightforward algebra finally
leads to 
\begin{eqnarray}
|\psi ^{(3)}\rangle  &=&\left( 1-T_{1}+T_{0}T_{1}-T_{1}T_{0}+\frac{1}{2}%
\tone^2-\frac{1}{2}T_{-1}T_{1}-\frac{1}{2}%
T_{-1}\tone^2-T_{1}\tzero^2-\tzero^2T_{1}+2T_{0}T_{1}T_{0}\right. 
\label{Psi3bis} \\
&&\left. -\frac{1}{2}T_{1}T_{0}T_{1}+\frac{3}{4}\tone^2T_{0}-\frac{1}{4}%
T_{0}\tone^2+T_{-1}T_{0}T_{1}-T_{-1}T_{1}T_{0}+\frac{3}{2}%
T_{1}T_{-1}T_{1}\right) |\varphi _{0}^{(3)}\rangle ,  \nonumber
\end{eqnarray}%
where $|\varphi _{0}^{(3)}\rangle $ is the orthonormal set of the $m=0$,
singly-occupied eigenstates. With the help of this form of $|\psi
^{(3)}\rangle $ the third-order expressions for the matrix element of an
operator $\langle O_{1}\rangle $ can be obtained (see Appendix \ref{app_A},
Eq. (\ref{O_3A})).%
\end{widetext}%

Rather cumbersome, but still straightforward calculations for the eigenvalue
problem with $|\psi ^{(3)}\rangle $ yield the fourth-order effective
Hamiltonian 
\begin{equation}
\mathcal{H}_{eff}^{(4)}=\mathcal{H}_{eff}^{(4),CT_{2}}  \label{Heff4}
\end{equation}%
%
%
%
%
that is \textit{identical} to the one obtained by the consecutive CT method,
Eqs.(\ref{H_CUT}). This effective Hamiltonian, within a unitary
transformation in the singly occupied subspace, is the $t$-$J$ model with
ring exchange and various correlated hoppings.\cite{Girvin:1989}


\subsubsection{Connection between eigenstates at different orders}

\label{EigenstatesLink}


There is a connection between eigenvectors in the $m=0$ subspace at
different orders in $t$. For definiteness, let us consider the eigenvalue
problem defined by $\mathcal{H}_{eff}^{(3)}$ in Eq. (\ref{Heff3}). If $%
|\varphi _{0}^{(2)}\rangle $ is an eigenstate of $\mathcal{H}_{eff}^{(3)}$,
then the corresponding energy to order $t^{3}$ is given by
\begin{equation}
\left( T_{0}-T_{-1}T_{1}+\emph{O}(t^{3})\right) |\varphi _{0}^{(2)}\rangle
=E|\varphi _{0}^{(2)}\rangle \ .
\end{equation}%
We wish to rewrite this equation as
\begin{equation}
\left( T_{0}-T_{-1}T_{1}\right) |\varphi _{0}^{(2)}\rangle =\left( E-\emph{O}%
(t^{3})\right) |\varphi _{0}^{(2)}\rangle .  \label{Conn1}
\end{equation}%
Note that the left-hand side of this expression contains $T_{0}-T_{-1}T_{1}=%
\mathcal{H}_{eff}^{(2)}$, the effective Hamiltonian at the previous order.
The eigenvalue problem for $\mathcal{H}_{eff}^{(2)}$ gives eigenstates at
the previous order, $|\varphi _{0}^{(1)}\rangle $ ($\equiv |\psi
_{0}^{(1)}\rangle $ in Eq. (\ref{psi1})). Taking the hermitian product of
Eq. (\ref{Conn1}) with an eigenstate $\langle \varphi _{0}^{\prime (1)}|$ we
have, by applying $T_{0}-T_{-1}T_{1}$ on the bra,
\begin{eqnarray}
\langle \varphi _{0}^{\prime (1)}|\left( T_{0}-T_{-1}T_{1}\right) |\varphi
_{0}^{(2)}\rangle &=&E^{\prime }\langle \varphi _{0}^{\prime (1)}|\varphi
_{0}^{(2)}\rangle \\
&=&\left( E-\emph{O}(t^{3})\right) \langle \varphi _{0}^{\prime (1)}|\varphi
_{0}^{(2)}\rangle .  \nonumber
\end{eqnarray}%
From this, one concludes that to order $t^{2}$,
\begin{equation}
\left( E-E^{\prime }\right) \langle \varphi _{0}^{\prime (1)}|\varphi
_{0}^{(2)}\rangle =0
\end{equation}%
Hence, to that order, either $\langle \varphi _{0}^{\prime (1)}|\varphi
_{0}^{(2)}\rangle =0$ or $E^{\prime }=E$ or both. The generalization of this
result means that an eigenvector in the low-energy subspace can have a
non-zero overlap to order $t^{n}$ with an eigenvector of the order $t^{n-1}$
theory if and only if the energies agree to order $t^{n-1}$. Note, by the
way, that the diagonalization of $\mathcal{H}_{eff}^{(n)}$ will in general
give us energies that contain all powers of $t$. Nevertheless, $E$ will be
valid only to order $t^{n}$ since the higher orders can be modified when the
matrix elements of $\mathcal{H}_{eff}$ are calculated to higher order. A
degeneracy that exists at a given order in $t$ can be lifted at the next
order. Energy levels of different symmetry can cross when evaluated at
different orders in $t$ so that the ground state of, for example, the $t-J$
model with correlated hopping 
can be different from that of the model that includes ring exchange.


\section{Resolvent method}

\label{resolvent}\setcounter{subsubsection}{0}

Another approach to deriving the low-energy effective theory is the
resolvent method. It is based on an iterative execution of a procedure known
as Lowdin downfolding.\cite{Lowdin} It bears a lot of similarity with BW
perturbation theory but works more directly with the Hamiltonian matrix
rather than with the eigenstates. Of all the approaches considered in this
paper the resolvent method requires the least amount of algebra.

We start by defining the projection operators $P_{m}$, $m=0,1,2$ and $P_{>}$%
. Operator $P_{m}$ projects on a subspace with $m$ doubly occupied sites.
Operator $P_{>}$ projects on a subspace with more than two doubly-occupied
sites
\begin{equation}
P_{>}=1-\sum_{m=0}^{2}P_{m}.
\end{equation}%
For our purposes it is convenient to rewrite Eq. (\ref{expansion_psi_m}) as
\begin{eqnarray}
&&|\psi \rangle =|\psi _{0}\rangle +|\psi _{1}\rangle +|\psi _{2}\rangle
+|\psi _{>}\rangle ,  \nonumber \\
&&P_{i}|\psi _{j}\rangle =\delta _{ij}|\psi _{j}\rangle \ ,
\end{eqnarray}%
where $i,j=0,1,2,$ and ``$>$''. The eigenvector equation can then be written in the
following block form
\begin{equation}
\left( 
\begin{tabular}{llll}
$T_{0}$ & $\;\;T_{-1}$ &  &  \\ 
$T_{1}$ & $1+T_{0}$ & $\;\;T_{-1}$ &  \\ 
& $\;\;T_{1}$ & $2+T_{0}$ & $T_{-1}$ \\ 
&  & $\;\;\;T_{1}$ & $\mathcal{H}_{>}$%
\end{tabular}%
\right) \left( 
\begin{tabular}{l}
$|\psi _{0}\rangle $ \\ 
$|\psi _{1}\rangle $ \\ 
$|\psi _{2}\rangle $ \\ 
$|\psi _{>}\rangle $%
\end{tabular}%
\right) =E\left( 
\begin{tabular}{l}
$|\psi _{0}\rangle $ \\ 
$|\psi _{1}\rangle $ \\ 
$|\psi _{2}\rangle $ \\ 
$|\psi _{>}\rangle $%
\end{tabular}%
\right) ,  \label{RM}
\end{equation}%
where $\mathcal{H}_{>}=P_{>}\mathcal{H}P_{>}$. Similarly to Eq. (\ref{BW})
in the BW formalism one needs to keep only terms up to $|\psi _{2}\rangle $
to derive the effective theory to order $t^{4}$. Thus, Eq. (\ref{RM}) should
suffice for our goals. It is convenient to rewrite this equation by
components 
\begin{eqnarray}
&&E|\psi _{0}\rangle =T_{0}|\psi _{0}\rangle +T_{-1}|\psi _{1}\rangle ,
\label{psi0r} \\
&&E|\psi _{1}\rangle =(1+T_{0})|\psi _{1}\rangle +T_{1}|\psi _{0}\rangle
+T_{-1}|\psi _{2}\rangle ,  \label{psi1r} \\
&&E|\psi _{2}\rangle =(2+T_{0})|\psi _{2}\rangle +T_{1}|\psi _{1}\rangle
+T_{-1}|\psi _{>}\rangle ,  \label{psi2r} \\
&&E|\psi _{>}\rangle =\mathcal{H}_{>}|\psi _{>}\rangle +T_{1}|\psi
_{2}\rangle .  \label{psi>r}
\end{eqnarray}%
Now we eliminate all components of $|\psi \rangle $ one by one, starting
with $|\psi _{>}\rangle $ until only $|\psi _{0}\rangle $ is left. From Eq. (%
\ref{psi>r}) we obtain 
\[
|\psi _{>}\rangle =\left( E-\mathcal{H}_{>}\right) ^{-1}T_{1}|\psi
_{2}\rangle =O(t)|\psi _{2}\rangle , 
\]%
where we take into account the fact that the operator in brackets is
non-singular and is $O(1)$. This expression for $|\psi _{>}\rangle $ is
substituted in the equation (\ref{psi2r}) for $|\psi _{2}\rangle $ to give 
\begin{widetext}%
\begin{eqnarray}
|\psi _{2}\rangle =\big(E-2-T_{0}-O(t)\big)^{-1}T_{1}|\psi _{1}\rangle
 =\left( -\frac{1}{2}T_{1}+O(t^{2})\right) |\psi _{1}\rangle , \label{psi2r_exp}
\end{eqnarray}%
where we expanded the denominator and kept terms of order $t$ since the
higher order terms do not contribute to the theory of the required $t^{4}$
order. This latter equation is used to eliminate $|\psi _{2}\rangle $ from
Eq. (\ref{psi1r}). Thus we  have
\begin{eqnarray}
|\psi _{1}\rangle =\left( -T_{1}-(E-T_{0})T_{1}-\frac{1}{2}%
T_{-1}\tone^2-(E-T_{0})^{2}T_{1}+O(t^{4})\right) |\psi
_{0}\rangle ,   \label{psi1r_exp} 
\end{eqnarray}%
where again the expansion of a denominator has been performed to the
required order. Finally, we obtain an equation for $|\psi
_{0}\rangle $ 
\begin{eqnarray}
E|\psi _{0}\rangle &=&\left( T_{0}-T_{-1}T_{1}-T_{-1}ET_{1}+T_{-1}T_{0}T_{1}-%
\frac{1}{2}\tm^2\tone^2\right.  \nonumber \\
&&\left. \vphantom{\frac{1}{2}}%
-T_{-1}E^{2}T_{1}-T_{-1}\tzero^2T_{1}+2ET_{-1}T_{0}T_{1}+O(t^{5})\right)
|\psi _{0}\rangle .
\end{eqnarray}%
This is not the \textquotedblleft true\textquotedblright\ eigenvalue
equation since it contains $E$-dependent terms in the right-hand side,
similar to the BW case. We rewrite it then by transferring $E$-dependent
terms to the left as 
\begin{equation}
E\big(1+T_{-1}T_{1}+ET_{-1}T_{1}-2T_{-1}T_{0}T_{1}\big)|\psi _{0}\rangle
=\left( T_{0}-T_{-1}T_{1}+T_{-1}T_{0}T_{1}-\frac{1}{2}%
\tm^2\tone^2-T_{-1}\tzero^2T_{1}\right) |\psi _{0}\rangle ,
\label{ER1}
\end{equation}%
where we omitted $t^{5}$-order terms. Below we will assume that operator
expression accuracy is up to $O(t^{4})$.

The derived equation for $|\psi _{0}\rangle $ still does not have the form
of the Schr\"{o}dinger equation $E|\psi \rangle =\mathcal{H}|\psi \rangle $.
To transform (\ref{ER1}) into a Schr\"{o}dinger equation an
orthogonalization procedure similar to the one we used in BW calculations
must be performed. We introduce $|\chi _{0}\rangle $ such as
\begin{eqnarray*}
|\psi _{0}\rangle =\alpha |\chi _{0}\rangle
\equiv \bigg(1+T_{-1}T_{1}+ET_{-1}T_{1}-2T_{-1}T_{0}T_{1}\bigg)%
^{-1/2}|\chi _{0}\rangle .
\end{eqnarray*}%
Substituting in Eq. (\ref{ER1}), left multiplying by $\alpha $ and then
expanding the square root keeping third order terms, the left-hand side is $%
E|\chi _{0}\rangle $ and the right-hand side still has some $E$-dependent
terms. Transferring them to the left we find that $|\chi _{0}\rangle $
satisfies
\begin{eqnarray}
&&E\left( 1+\frac{1}{2}\left( T_{-1}T_{1}T_{0}+T_{0}T_{-1}T_{1}\right)
\right) |\chi _{0}\rangle =\left( T_{0}-T_{-1}T_{1}-\frac{1}{2}%
T_{-1}T_{1}T_{0}-\frac{1}{2}T_{0}T_{-1}T_{1}+T_{-1}T_{0}T_{1}\right. \\
&&\phantom{E 1 + \frac{1}{2} T_{-1} T_1 T_0}\left.
+T_{-1}T_{0}T_{1}T_{0}+T_{0}T_{-1}T_{0}T_{1}+(T_{-1}T_{1})^2-\frac{1}{%
2}\tm^2\tone^2-T_{-1}\tzero^2T_{1}\right) |\chi _{0}\rangle .
\nonumber
\end{eqnarray}%
The left-hand side of this equation still does not have the desired form. An
extra orthogonality transformation analogous to that performed with $\alpha $
above is required
\[
|\chi _{0}\rangle =\left( 1+\frac{1}{2}T_{-1}T_{1}T_{0}+\frac{1}{2}%
T_{0}T_{-1}T_{1}\right) ^{-1/2}|\varphi _{0}\rangle ,
\]%
with the subsequent square root expansion. The resulting eigenvalue equation
for $|\varphi _{0}\rangle $ by the resolvent method finally takes the Schr%
\"{o}dinger equation form $E|\varphi _{0}\rangle =\mathcal{H}%
_{eff}^{(4)R}|\varphi _{0}\rangle $ with the effective Hamiltonian given, to
fourth order, by
\begin{equation}
\mathcal{H}_{eff}^{(4)R}=\mathcal{H}_{eff}^{(4)CT_{1}}+\frac{1}{4}%
(T_{-1}T_{1}\tzero^2+\tzero^2
T_{-1}T_{1})-\frac{1}{2}%
T_{0}T_{-1}T_{1}T_{0}\ .  \label{HR}
\end{equation}%
\end{widetext}%
This effective Hamiltonian does not coincide with the other forms of $%
\mathcal{H}_{eff}^{(4)}$ we have obtained so far, namely
Eqs. (\ref{AM}), (\ref%
{H_CUT}), and (\ref{Heff4}). It is however unitarily related to all others
through the transformation Eqs. (\ref{Canonical_T}), (\ref{Canonical_T_explicit}%
). For example, the \textquotedblleft angle of rotation\textquotedblright\ $%
\gamma =1/4$ transforms Eq. (\ref{HR})
back to $\mathcal{H}_{eff}^{(4),CT_{1}}$ in Eq. (\ref{AM}).

We note that transformation of operators can also be devised within the
resolvent approach in a manner similar to the BW calculations, Sec. \ref{DBW}%
, using the above relations between $|\varphi _{0}\rangle $ and the states
with $m> 0$ doubly-occupied sites.


\section{Big Picture: General Transformation}

\label{bigp}


Thus far, we have shown that while different methods of performing
 perturbation theory preserve the original energy spectrum of the
 Hubbard model, the effective 
 Hamiltonians, Eqs. (\ref{AM}), and Eq. (\ref{HR}) differ from each
 other and from  Eqs. (\ref%
{H_CUT}), (\ref{Heff4}) that agree with each other.  The terms in
 question all arise 
at fourth order and are all reducible\cite{pgs} with respect to the
 zero-double occupancy 
sector.  That is, they contain hopping processes that do not originate
 from excitation to the doubly occupied subspace, $T_0T_{-1}T_1T_0$,
 nor terminate once an electron is returned to the 
 singly occupied subspace, for example, $T_{-1}T_1T_{-1}T_1$. All such
 processes can be viewed as arising from a
 transformation\cite{Stein} of the
 eigenstates in the low-energy sector.  To lay plain how the effective
 Hamiltonian is unavoidably affected by this transformation, we now
 formulate a general method which makes it possible to derive all of
 the Hamiltonians presented thus far within a single computation
 scheme.   Our starting point is BW integral equation,
 Eq. (\ref{Projection}), whose solution we write symbolically as 
\beq\label{gdef}
|\psi\rangle=G(t,E)|\psi_0\rangle.
\eeq
The exact expression for the energy-dependent operator $G(t,E)$ is obtained by iterating Eq. (\ref{Projection}). 
 Applying $P=1-Q$ (see Sec. \ref{DBW}) to the left-hand side of the Schr\"odinger equation, Eq. (\ref{Schroedinger}), we obtain
\beq
(E-T_0)|\psi_0\rangle-PT_{-1}Q|\psi\rangle=0 ,
\eeq
which can be recast as a non-linear eigenvalue problem
\beq\label{epsi}
E|\psi_0\rangle=(T_0+PT_{-1}G(t,E))|\psi_0\rangle ,
\eeq
using Eq. (\ref{gdef}). Taylor expansion of $G(t,E)$ results in a polynomial in the energy eigenvalue.  Through fourth order we find that,
\begin{widetext}
\beq\label{epsi1}
E|\psi_0\rangle=\left(
T_0-T_{-1}T_1+T_{-1}T_0T_1-T_{-1}T_0^2T_1-\frac12T_{-1}^2T_1^2 
+\bigg(-ET_{-1}T_1-E^2T_{-1}T_1+2ET_{-1}T_0T_1\bigg)\right)|\psi_0\rangle.
\eeq
To eliminate the energy-dependence on the right-hand side of this
equation, we substitute Eq. (\ref{epsi}) for each occurrence of
$E|\psi_0\rangle$ until all the energy dependence has disappeared.
The result of this procedure is an eigenvalue problem
\beq
E|\psi_0\rangle&=&\left(T_0-T_{-1}T_1+T_{-1}T_0T_1-T_{-1}T_0^2T_1-\frac12T_{-1}^2T_1^2
-T_{-1}T_1(T_0-T_{-1}T_1)+T_{-1}T_1T_0^2+
2T_{-1}T_0T_1T_0\right)|\psi_0\rangle\nonumber\\  
&=&\widetilde{H}|\psi_0\rangle ,
\eeq
\end{widetext}
with a non-hermitian operator $\widetilde{H}$. All the terms in the
second parenthesis arise explicitly from eliminating the energy
dependence in Eq. (\ref{epsi1}) and as a consequence are products of
the proper BW terms in the first parenthesis. It is in this sense that
such terms are reducible with respect to the zero double occupancy
sector. The lack of hermiticity surfaces because projection does not
respect the mutual orthogonality of the eigenstates in any of the
degenerate subspaces.

Hermiticity can be restored by a suitable global transformation of the
eigenstates within each degeneracy subspace.  To proceed, we consider
the operator 
\beq\label{zeff}
Z=1+aT_{-1}T_1+bT_{-1}T_1T_0+cT_0T_{-1}T_1+dT_{-1}T_0T_1 , \nonumber
\eeq
which is explicitly not unitary.
 As such, it can be used to construct an effective Hamiltonian
\beq
{\cal H}_{\rm eff}=Z^{-1}\widetilde{H}Z
\eeq
by placing appropriate conditions on the coefficients, $a$, $b$, $c$, and $d$
so that ${\cal H}_{\rm eff}$ is hermitian.
Through fourth order, hermiticity is restored by demanding that
\beq
a&=&-\frac12\nonumber\\
b&=&\gamma-\frac12\\
c&=&-\gamma-\frac12\nonumber\\
d&=& 1\nonumber
\eeq
The resultant effective Hamiltonian,
\begin{widetext}
\beq
{\cal H}^{(4)}_{\rm eff}&=&\tzero-\tm\tone+\tm\tzero\tone-\frac12(\tm\tone\tzero+\tzero\tm\tone)
-\tm\tzero^2\tone-\frac12\tm^2\tone^2+\tm\tzero\tone\tzero+\tzero\tm\tzero\tone\nonumber\\
&+&(\tm\tone)^2-\frac12(\tm\tone\tzero^2+\tzero^2\tone\tm)
+\gamma(2\tzero\tm\tone\tzero-\tm\tone\tzero^2-\tzero^2\tm\tone)
\eeq
\end{widetext}
contains three reducible terms whose magnitude is set by an arbitrary
 constant $\gamma$.  These terms are given precisely by the ``additional''
canonical transformation in Eq. (\ref{Canonical_T_explicit}). 
 Because all the terms controlled by the magnitude
of $\gamma$ are reducible, they provide no more than a transformation of the eigenstates
within the degeneracy subspace with zero double occupancy.  The
 multiplicity of Hamiltonians we have derived here all arise from
 different choices for $\gamma$. 
For example, 
within the canonical transformation method of
 Ref. \onlinecite{Girvin:1989}, we have $\gamma=0$.  Effective
 Hamiltonians 
within a sector with a fixed number of doubly occupied sites can only
 be determined up to an arbitrary rotation of the eigenstates within
 the degeneracy space.  To understand what happens in the case of BW
 perturbation theory, we recall that, although one starts from a set
 of wave functions that are orthogonal in the full Hilbert space, the
 projection into a degeneracy subspace is a process that does not
 respect the orthogonality. To get a hermitian Hamiltonian we have to
 perform a general transformation on the degeneracy subspace in such a
 way that the orthogonality of the projected components is
 restored. This is what the general non-unitary transformation defined
 by Eq. (\ref{zeff}) is doing. Canonical transformations, on the other
 hand, are by definition unitary and hence no additional
 orthogonalization transformations are
 necessary.\cite{Eskes:1994,Harris:1967}

\section{Some comments on experiment}

\label{Discussion}

Several experimental groups \cite{Experiments} have pointed out that
high-temperature superconductors show spectral weight rearrangements over
the Mott scale.\cite{wrobel}   Rearrangement of spectral weight over large
energy scales is expected in strongly correlated systems simply because
many of the eigenstates are localized or almost localized. Using the Lehman
representation, one can easily see that the momentum eigenstates probed by
photoemission or optical spectroscopy, for example, have non-vanishing
projection on essentially all the true eigenstates of the interacting
problem. When the Mott gap is closed, this means that spectral weight
changes will occur over all the energy scales when doping or temperature is
changed. When the Mott gap is opened, this will continue to be the case but,
nevertheless, the spectral rearrangements over the lower Hubbard band will
be describable to a high degree of accuracy using only the effective
low-energy theory, \textit{as long as one uses the operators that are
appropriate for the low-energy sector.} These operators take into account
rearrangements in the upper Hubbard band through virtual states. For
example, upon doping by an amount $x$, exact calculations on the Hubbard
model show that the spectral weight transfered from the upper to the lower
Hubbard band exceeds $2x,$ while $2x$ was argued to be the
prediction of the $t-J$ model by some early work.\cite{Meinders:1993}
Even at the level of the $t-J$ model, however, there is a correction\cite{Harris:1967,Eskes:1994}
to the low-energy spectral weight (LESW) from the transitions across
the Hubbard gap that arise from transforming the electron
operators. Through order $(t/U)^2$, the LESW agrees well with the
exact diagonalization on small systems.


\section{Conclusion}

\label{Conclusions}

We have studied several methods for performing degenerate perturbation theory.
We have shown that, to fourth order, they lead to low-energy effective
theories that appear different but, in fact, are all
related through a unitary transformation in the low-energy subspace,
Eqs. (%
\ref{Canonical_T}), (\ref{Canonical_T_explicit}). The necessity of a unitary
transformation in the low-energy subspace to prove the equivalence of the
theories does not normally occur in lower-order theories
and thus is a rather unfamiliar property.
The most systematic approaches are the two canonical transformation methods,
the easiest algebraically are the general transformation (Sec. \ref{bigp})
and the resolvent method, while the
Brillouin-Wigner method, modified \`{a} la Rayleigh-Schr\"{o}dinger,
becomes rather cumbersome in higher orders.
Nevertheless, the latter method gives some
insight into the other approaches. In particular, it allows to understand
the appearance of terms in intermediate states that appear, at first glance,
to be in the low-energy subspace (\textquotedblleft improper
terms\textquotedblright ). Also, it shows that in the low-energy subspace,
eigenvectors at order $t^{n}$ have non-zero projection on eigenvectors at
order $t^{n-1}$ only if the expansions of the corresponding energies agree
to order $t^{n-1}$. The need to orthogonalize and to keep track of the order
of the energy expansion makes the BW approach in practice more delicate to
carry out to high order. However, the existence of a small parameter makes
it completely equivalent to the canonical transformation and resolvent
approaches.

This work has laid plain how perturbation
theory to all orders can be formulated unambiguously. 
Although the question of the accuracy of
projected schemes in treating the full Hubbard model has not been
addressed in the present work,
 it is generally assumed 
that such projected schemes should capture all the physics of the
original Hubbard model to a given order in $t/U$. Note, however, that
this presupposes that in the strong coupling limit all observables
 can be expanded in powers of $t/U$. For an infinite system, this may
 not always be true.
 Also, as is the case with any
 perturbation theory, such projected schemes obviously fail
 to adequately describe phase transitions. For example, at
 half-filling the interaction tuned Mott transitions is beyond the
 domain of applicability of the projected theory. In the case of the
 doping-induced Mott transition, the projected theory cannot address
 the question of the chemical potential, which is of order $U$. 
 Nonetheless, as long as the interaction $U$ is much larger than the
bandwidth, it is possible to write a unique (up to a unitary
transformation in the target space) low-energy effective theory whose
range of applicability is limited by the condition that physical
observables have a well-defined expansion in $t/U$.  We reiterate that
consistency of such procedures requires that all operators be
transformed as well. 
Although we presented the results within the Hubbard model, all  methods
are easily generalizable to other models.


\begin{acknowledgments}
A.-M.S.T. would like to thank A. Blais, J.-Y. Delannoy, M. Gingras,
and S. Girvin for useful conversations. A.L.C. acknowledges useful
conversations with S. R. White and A. H. Castro Neto.
The present work was supported, in part, by a Research Corporation Award
(A.L.C), by NSF under Grant No. DMR-030586 (D.G. and P.P.), and
by NSF Grant No. 0342157 at Yale and by the Natural Sciences and
Engineering Research Council (NSERC) of Canada, the Canadian Institute for
Advanced Research, and the Tier I Canada Research Chair Program
(A.-M.S.T.).
\end{acknowledgments}


\appendix

\section{Miscellaneous additional results}

\label{app_A}

The third-order CT generator $S_3$, version of Ref. \onlinecite{Girvin:1989}%
, is given by
\begin{eqnarray}  \label{S3}
S_3&=&\frac23\bigg( \tm^2T_1+T_1\tm^2-T_{-1}T_{1}^{2}  \nonumber
\\
&&-T_{1}^{2}T_{-1}+2T_1T_{-1}T_1-2T_{-1}T_1T_{-1}\bigg) \\
&&+T_{0}^{2}T_{-1}+T_{-1}T_{0}^{2}-T_1T_{0}^{2}-T_{0}^{2}T_1  \nonumber \\
&&-2T_0T_1T_0+2T_0T_{-1}T_0.  \nonumber
\end{eqnarray}

The BW third-order expressions for the matrix element of an off-diagonal
operator $\langle O_{1}\rangle $ increasing the number of doubly-occupied
sites by one is given by
\begin{eqnarray}  \label{O_3A}
&&\langle \psi^{\prime(3)}\vert O_{1}\vert \psi^{(3)}\rangle  \nonumber \\
&&\phantom{\langle \psi^{\prime(3)}|} =\langle \varphi_{0}^{\prime(3)}\vert
- T_{-1}O_1 +(T_{-1}T_0-T_0T_{-1})O_1  \nonumber \\
&&\phantom{\langle \psi^{\prime(3)}|} +\frac32 T_{-1}T_1T_{-1}O_1-\frac12
\tm^2T_1O_1 \\
&&\phantom{\langle \psi^{\prime(3)}|}
-T_{0}^{2}T_{-1}O_1-T_{-1}T_{0}^{2}O_1+2T_0T_{-1}T_0O_1  \nonumber \\
&&\phantom{\langle \psi^{\prime(3)}|} +\frac12 T_{-1}O_1T_{-1}T_1-\frac12
\tm^2O_1T_1\vert \varphi_{0}^{(3)}\rangle .  \nonumber  \label{O_3}
\end{eqnarray}


\end{document}